\let\origvec\vec
\documentclass[runningheads]{llncs}
\usepackage[T1]{fontenc}
\usepackage{graphicx}
\usepackage[linesnumbered,ruled,vlined]{algorithm2e}
\usepackage{amssymb}

\let\vec\origvec
\usepackage{amsmath}
\usepackage[
    style=ieee,
    doi=false,
    isbn=false,
    url=false,
    eprint=false,
    backend=biber,
    natbib=true
    ]{biblatex}
\bibliography{references}

\usepackage[colorlinks=true,linkcolor=blue,citecolor=blue]{hyperref}
\usepackage{color}

\usepackage{cleveref}
\newtheorem{rem}{Remark}
\begin{document}
\newcommand{\macro}[1]{\texttt{\textbackslash#1}}


\newcommand{\R}{\mathbb{R}}
\newcommand{\N}{\mathbb{N}}
\newcommand{\SO}{\mathsf{SO}}
\newcommand{\SE}{\mathsf{SE}}

\newcommand{\regtext}[1]{\mathrm{\textnormal{#1}}}
\newcommand{\defemph}[1]{\textbf{#1}}
\newcommand{\st}[1]{_{\regtext{#1}}}
\newcommand{\mc}[1]{\mathcal{#1}}

\newcommand{\niceinf}[2]{\inf_{#1}~#2}

\newcommand{\emptyarr}{[\ ]}
\newcommand{\zeros}{\mathbf{0}}
\newcommand{\ones}{\mathbf{1}}
\newcommand{\eye}{\mathbf{I}}

\newcommand{\inv}{^{-1}}
\newcommand{\comp}{^{\regtext{C}}}
\newcommand{\norm}[1]{\left\Vert#1\right\Vert}
\newcommand{\trans}{^{\top}}

\newcommand{\nom}{_{\regtext{nom}}}

\newcommand{\state}{x}
\newcommand{\statetranspose}{x^{\top}}
\newcommand{\stateQ}{Q^i_t}
\newcommand{\stateq}{q^i_t}
\newcommand{\control}{u}
\newcommand{\controlR}{R^{ij}_{t}}
\newcommand{\controlr}{r^{ij}_{t}}
\newcommand{\Rii}{R^{ii}_{t}}
\newcommand{\rii}{r^{ii}_{t}}
\newcommand{\Riiinv}{R^{ii^{-1}}_{t}}
\newcommand{\Rij}{R^{ij}_{t}}
\newcommand{\rij}{r^{ij}_{t}}
\newcommand{\Rjjinv}{R^{jj^{-1}}_{t}}
\newcommand{\rjj}{r^{jj}_{t}}

\newcommand{\dynA}{A_{t}}
\newcommand{\dynB}{B^{j}_{t}}
\newcommand{\dynBtranspose}{B^{j\top}_{t}}





%
\title{Game-theoretic Occlusion-Aware Motion Planning:
an Efficient Hybrid-Information Approach}
%
%
\author{Kushagra Gupta\orcidID{0009-0005-8927-671X}\and
David Fridovich-Keil\orcidID{0000-0002-5866-6441}
}
\authorrunning{K. Gupta, D. Fridovich-Keil}
%
\institute{The University of Texas at Austin \\
\email{\{kushagrag, dfk\}@utexas.edu}\\
}
\maketitle              
\begin{abstract}

We present a novel algorithm for game-theoretic trajectory planning, tailored for settings in which agents can only observe one another in specific regions of the state space. Such problems arise naturally in the context of multi-robot navigation, where occlusions due to environment geometry naturally mask agents' view of one another. In this paper, we formalize these settings as dynamic games with a hybrid information structure, which interleaves so-called ``open-loop'' periods (in which agents cannot observe one another) with ``feedback'' periods (with full state observability). We present two main contributions. First, we study a canonical variant of these hybrid information games in which agents' dynamics are linear, and objectives are convex and quadratic. Here, we build upon classical solution methods for the open-loop and feedback variants of these games to derive an algorithm for the hybrid information case that matches the cubic runtime of the classical settings. Second, we consider a far broader class of problems in which agents' dynamics are nonlinear, and objectives are nonquadratic; we reduce these problems to sequences of hybrid information linear-quadratic games and empirically demonstrate that iteratively solving these simpler problems with the proposed algorithm yields reliable convergence to approximate Nash equilibria through simulation studies of overtaking and intersection traffic scenarios.

\keywords{Dynamic Game Theory  \and Multi-Agent Motion Planning.}

\end{abstract}
\section{Introduction}


This paper considers the setting of \(N\) agents playing a smooth, discrete-time, general-sum dynamic game with a \emph{hybrid} information structure: at any time step, either (i) every agent can see each other or (ii) agents do not have mutual visibility over one another. This setting is particularly important for representing occlusions in multi-agent motion planning problems commonly encountered in robotics. \par
Dynamic game theory provides a suitable formalism for motion planning problems involving multiple decision-making agents and avoids requiring an agent to make \emph{a priori} predictive assumptions about others \cite{lavalle1995game}. Dynamic games require the specification of an \emph{information structure}, i.e., the information available to each agent at each time $t$, in order for equilibrium strategies to be determined. The choice of information structure is critical because in a dynamic game, each agent has to consider their interactions with all other agents, and consequently, the information available to each agent during decision-making naturally influences the game solution. In this paper, we consider the two extremes on the spectrum of information structures -- open-loop and feedback.

Employing an open-loop structure implies that every agent will have no information about any other agents' state throughout the duration of the game except knowing their initial states. In short, in an open-loop game, agents must play while \emph{blindfolded}. On the other hand, employing a feedback structure implies that all agents have access to current state information about every other agent.

\begin{figure}
    \vspace*{-0.5cm}
    \centering
    \includegraphics[width=0.7\columnwidth]{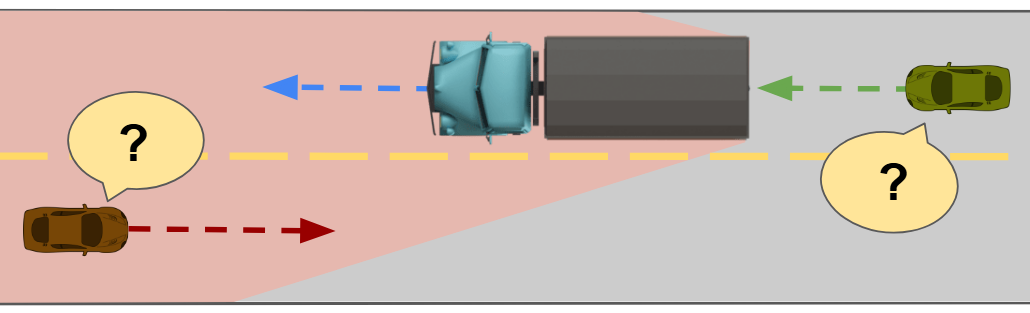}
    \caption{Example of an occlusion scenario potentially benefiting from a hybrid information dynamic game formulation. The green and brown cars cannot see each other currently, yet each car's current actions will influence the cars’ future visibility and, ultimately, one another’s future strategy.}
    \label{intro_fig}
    \vspace*{-0.5cm} 
\end{figure}

Existing dynamic game formulations usually assume that only one of the feedback or open-loop structures holds throughout the game. However, in the context of motion planning, using only one of the information structures is often inadequate to accurately represent real-life interaction scenarios, such as the presence of \emph{occlusions} throughout a portion of the planning horizon. For example, consider Fig. \ref{intro_fig}, in which a large blue truck is moving on a two-way expressway. A green car driving in the same direction approaches this slow-moving truck and wishes to overtake it. However, the green car is unable to see the brown car travelling in the opposite direction while it is just behind the truck due to an occlusion made by the (physically larger) truck. At this moment, the green car needs to decide how to overtake the truck. Similarly, the brown car has to decide whether to deviate from its current path when it sees the green car overtaking the blue truck. This example highlights a core challenge in game-theoretic motion planning: the ability of autonomous agents to \emph{predict} or \emph{reason} about behaviours of other agents is compromised in the presence of occlusions. Currently, no existing information structure can addresses this challenge adequately. During an occlusion, using a feedback structure is inappropriate. On the other hand, though it is possible to use an open-loop structure throughout the game, this choice does not allow agents to exploit knowledge of one another’s state when occlusions are absent, which will naturally offer an opportunity for better decision-making.

The above example motivates the development of a \emph{hybrid} information structure, which interleaves both open-loop and feedback structures in the same dynamic game. When occlusions are present, adopting an open-loop structure to model agent interactions is an attractive option: the agents can play an open-loop game whilst having no visibility of other agents. Similarly, a feedback structure can be adopted at other times when occlusions are absent, and agents can see one another.
\\
\textbf{Contributions.} Although analytical solutions for open-loop and feedback games with linear dynamics and convex, quadratic costs (LQ) exist, these closed-form expressions cannot be directly interleaved to make a hybrid information structure as described above. To this end, we propose two contributions in this paper. 
First, we introduce an algorithm to find Nash equilibria in an LQ game with a hybrid information structure played under known occlusion patterns. 
 Second, we develop an iterative framework, \textbf{O}cclusion-aware \textbf{G}ame \textbf{Solve}r (\texttt{OGSolve}), which is applicable to a broader class of problems where agents have nonlinear dynamics and nonquadratic costs. \texttt{OGSolve} finds and solves a sequence of hybrid information LQ games, yielding local approximations of Nash equilibria in non-LQ games with a priori unknown occlusion patterns.

\vspace{-0.3cm}
\section{Related Work}\label{Related Work}
We begin by highlighting advances in game-theoretic motion planning. We then discuss iterative linear quadratic methods to provide context for our overall framework. \\
\textbf{Game Theory in Motion Planning.} Motion planning is a well-studied problem in the robotics community, and an important sub-problem is planning in environments containing occlusions. Since the first game-theoretic approaches for motion planning \cite{lavalle1993game}, early work addressing occlusions focused on maintaining visibility to minimize uncertainty caused by occlusions by formulating the problem as a pursuit-evasion game \cite{lavalle1999visibility, vidal2002probabilistic} or a target-tracking game \cite{lavalle1997motion}, using zero-sum differential games. The general multi-agent motion planning problem has attracted the adoption of non-cooperative game-theoretic optimal control approaches, including feedback differential game \cite{lavalle1998optimal, jha2015game} and open-loop game formulations \cite{zhu2014game}.\\
\textbf{Hybrid Information Games for Occlusion-Aware Planning.} Methods for multi-agent occlusion-aware planning combining feedback and open-loop approaches are not yet well studied, but the mixture of these information structures offers a promising framework for occlusion-aware motion planning due to their convenient occlusion modelling potential. In particular, a method adopting such a hybrid approach is proposed by Zhang et al. \cite{Zhan-RSS-21}, involving a hybrid zero-sum dynamic game used to plan safe trajectories under occlusions. Their work assumes only two adversarial agents. However, occlusion scenarios, in general, can include \(N>2\) non-adversarial agents. Further, their approach has exponential complexity and is incapable of real-time performance. In our work, we use an efficient iterative framework to seek Nash equilibrium approximations in hybrid information general-sum games with \(N\) agents, achieving cubic complexity.\\
\textbf{Iterative LQ Methods.} A fundamental problem in optimal control is that of the Linear Quadratic Regulator (LQR) \cite{kalman1960contributions, kalman1960new}, which is the basis of many iterative trajectory optimization methods that yield a sequence of states and controls that minimize a given cost function locally \cite{mayne1966second, jacobson1970differential}. Such methods iteratively refine a trajectory and can accommodate nonlinear dynamics and non-quadratic costs by taking local quadratic approximations of costs and local linear approximations of dynamics and then solving the corresponding LQ subproblem \cite{li2004iterative, tassa2014control, todorov2005generalized}.
In doing so, they find a local solution and avoid the curse of dimensionality associated with seeking global solutions in high dimensions \cite{bellman1959mathematical}. 
Motion planning problems involving multiple decision-making agents also have LQ problem formulations. In particular, our work employs the ILQGames \cite{fridovich2020efficient} framework, which finds real-time local solutions to $N$-player general-sum differential games.
\vspace{-0.3cm}
\section{Preliminaries}\label{Preliminaries}

In this section, we briefly introduce concepts underlying our approach. We start by giving an overview of Nash equilibrium solutions to linear-quadratic discrete time dynamic games for both open-loop and feedback information structures. We then describe an iterative approach that we shall use to extend the scope of our work beyond games with linear dynamics and quadratic costs.
\vspace{-0.5cm}
\subsubsection{Nash Equilibrium solutions to Linear Quadratic (LQ) games.}\label{3A}
Consider an $N$-person discrete-time, deterministic, infinite dynamic game of a fixed duration of $T$ time steps, with dynamics given by \(x_{t+1} = f_{t}(x_{t}, u^{1}_{t},..., u^{N}_{t})\), \(t\in \{1,...,T\} \equiv \mathbf{T}\)
where \(x_{t}=(x^1_{t},\dots,x^N_{t}) \in \mathbb{R}^{n}\) denotes the concatenated states of all players at time $t$. The initial state $x_{1}$ is given \emph{a priori}; \(u^{i}_{t} \in \mathbb{R}^{m_{i}}\) denotes the control input of player $i$ at time $t$, 
$\forall i\in \{1,...,N\} \equiv \mathbf{N}$. In addition, each player has a time-additive cost function \(J^{i}(u^{1},..., u^{N}) = \sum_{t=1}^{T} g^{i}_t(x_{t}, u^{1}_{t},..., u^{N}_{t})\), \(i\in \mathbf{N}\) 
which it seeks to minimize (for a complete definition, see \cite[Chapter~5]{bacsar1998dynamic}). 

For both open-loop and feedback information structures discussed below, we consider a game with convex quadratic costs and linear dynamics
\vspace{-0.3cm}
\begin{equation}\label{lqcost}
    g^{i}_t(x_{t}, u^{1:N}_{t}) = \frac{1}{2} \Bigg[ \Big( \statetranspose_{t}Q^{i}_{t}+2q^{i\top}_{t} \Big) \state_{t}\, + \left. \sum_{i=1}^{N}\left(\control^{j\top}_{t}R^{ij}_{t} + 2r^{ij\top}_{t}\right)\control^{j}_{t}\right],
\end{equation}
\vspace{-0.3cm}
\begin{equation}\label{lindyn}
    f_t(x_t, u^{1:N}_{t})= \dynA\state_{t} + \sum_{j=1}^{N}\dynB\control^{j}_{t},
\end{equation}
where $u_t^{1:N}$ denotes the control input of all $N$ players at time $t$. \(Q^i_t\succeq0, q^i_t\) and \(R^{ij}_t\succeq0, r^{ij}_t (j\in\mathbf{N})\) denote the state and control cost matrices for the \(i^{\mathrm{th}}\) player at time \(t\). \(A_t\) and \(B_t^{i}, i\in\mathbf{N}\) represent the dynamics. We now provide a brief overview of open-loop and feedback solutions to LQ games and highlight the key sets of equations we use in our work in Algorithms \eqref{algo: solveLQOpenLoop}-\eqref{algo: solveLQFeedback}. For detailed derivations, see \cite[Chapter~6]{bacsar1998dynamic}.\\
\textbf{Open-Loop Nash Equilibria.} Open-loop games have an information structure where each player only knows the initial states of all other players. Thus for \(t\geq1, u^{i}_{t} \equiv \gamma^{i}_{t}(x_{1})\), \(i \in \mathbf{N}\), for some choice of strategy $\gamma^{i}_t:\mathbb{R}^n\rightarrow\mathbb{R}^{m_{i}}$.
To solve an open-loop LQ game, we consider the Lagrangian for each player \(i\)
\begin{equation}\label{eq: OL Lagrangian}
    \mathcal{L}^i(x_{1:T}, u_{1:T}^{1:N}, \lambda_{1:T}^{1:N}) = \sum_{t=1}^T g^{i}_t(x_{t}, u^{1:N}_{t}) - \sum_{t=1}^T \lambda^{i\top}_t\left(x_{t+1} - \dynA\state_{t} - \sum_{j=1}^{N}\dynB\control^{j}_{t}\right),
\end{equation}
where \(\lambda^{i}_t\) denotes the Lagrange multiplier for Player \(i\) at time \(t\). The Lagrangian \eqref{eq: OL Lagrangian} is used to solve the KKT conditions \cite{nocedal1999numerical}, which yields the following recursions that are solved backwards in time
to compute unique open-loop equilibrium controls
\vspace{-0.4cm}
\begin{equation}\label{reclambda}
\begin{split}
        \Lambda_{t} &= I + \sum_{j=1}^{N} \dynB \Rjjinv \dynBtranspose M^{j}_{t+1}, \\
     m^{i}_{t} &=\dynA^{\top}\Bigg[m^{i}_{t+1} - M^{i}_{t+1}\Lambda_{t}^{-1}\sum_{j=1}^{N}\dynB\Rjjinv \left(\dynBtranspose m^{j}_{t+1} \,+ \right.\left.\rjj\right)\Bigg]+q^{i}_{t},\\
     M^{i}_{t} &= \stateQ + \dynA M^{i}_{t+1}\Lambda^{-1}_{t}\dynA, \quad M^{i}_{T} = Q^{i}_{T}, \quad m^{i}_{T} = q^{i}_{T},\\
\end{split}
\end{equation}
with optimal open-loop equilibrium control and state values given by
\vspace{-0.3cm}
\begin{equation}\label{openloopcontrols}
    \begin{split}
     u^{i*}_{t} &= -\Riiinv\left[\dynBtranspose\left(M^{i}_{t+1}x^{*}_{t+1}+m^{i}_{t+1}\right) + \rii\right],\\
     x^{*}_{t+1} &= \Lambda^{-1}_{t}\left[\dynA x^{*}_{t} -\sum_{j=1}^{N} \dynB (R^{jj}_t)^{-1} \left( \dynBtranspose m^{j}_{t+1} + \rjj\right)\right].\\
    \end{split}
\end{equation}
\vspace{-0.5cm}
\begin{rem}\label{remark: OL}
Solving an open-loop LQ game requires \(\{M^{i}_{T},m^{i}_{T}\}\) as inputs, and finds \(\{M^{i}_{t}, m^i_t\}_{t\in\mathbf{T}}~\forall~i \in \mathbf{N}\) by using \eqref{reclambda}, as in Algorithm \ref{algo: solveLQOpenLoop}. The optimal open-loop controls can then be found for some initial state \(x_1\) by using \eqref{openloopcontrols}.
\end{rem}
\begin{algorithm}\label{algo: solveLQOpenLoop}
\SetKwComment{Comment}{/* }{ */}
\caption{Open-Loop LQ Game Solver (\texttt{solveLQOpenloop})}
\KwIn{time horizon $T$, linear dynamics (\ref{lindyn}), costs (\ref{lqcost}), \(M^i_T, m^i_T~\forall~i\in\mathbf{N}\)}
\KwOut{\(\{M^{i}_{t}, m^i_t\}_{t\in\mathbf{T}}~\forall~i \in \mathbf{N}\)}
\For{time $t=T,\dots,1$}{
compute recursions given in \eqref{reclambda} \hspace{2.4cm}\Comment{backward in time}
}
\KwRet{\(\{M^{i}_{t}, m^i_t\}_{t\in\mathbf{T}}~\forall~i \in \mathbf{N}\)}
\end{algorithm}
\vspace{-0.2cm}
\noindent
\textbf{Feedback Nash Equilibria.}\label{FN}
At any stage in a feedback game, each player has \emph{current} information about all other players. Thus every player can take the current state into account to determine their control strategy, i.e., for \(t\geq 1, u^{i}_{t} \equiv \gamma^{i}_{t}(x_{t})\), for some choice of strategies $\gamma^{i}_t:\mathbb{R}^n\rightarrow\mathbb{R}^{m_{i}}$.
The coupled Hamiltonian-Jacobi equations relate all players' value functions $V^{i}_{t}(x_{t})$, which measures their optimal cost to go as (with $V^{i}_{T+1}(x_{T+1})=0 \: \forall \, i \in \mathbf{N}$)
\vspace{-0.3cm}
\begin{equation}\label{valuefunc}
    V^{i}_{t}(x_{t}) = \min_{u^{i}_{t}}\bigg\{\frac{1}{2}\Big(\left( \statetranspose_{t}Q^{i}_{t}+2q^{i\top}_{t}\right)\state_{t} + \sum_{i=1}^{N}\left(\control^{j\top}_{t}R^{ij}_{t} + 2r^{ij\top}_{t}\right)\control^{j}_{t}\Big) + V^{i}_{t+1}(x_{t+1}) \bigg\}. 
\end{equation}
\vspace{-0.3cm}
It can be shown that the value function is quadratic, i.e.,
\begin{equation}\label{quadvalue}
    V^{i}_{t}(x_{t}) = \frac{1}{2}\left( x_{t}^{\top}Z^{i}_{t} + 2\zeta^{i\top}_{t}\right)x_{t} + n^{i}_{t},
    \vspace*{-0.3cm}
\end{equation}
with $Z^{i}_{T+1} = 0$, $\zeta^{i}_{T+1} = 0$ and $n^{i}_{T+1} = 0$. The feedback control law $\gamma^i_t$ is determined by finding the minimizer of \eqref{valuefunc} and substituting $x_{t+1}$ from \eqref{lindyn} inside \eqref{quadvalue}. We assume strong convexity and set the gradient (with respect to control) to zero, obtaining
\vspace{-0.3cm}
\begin{equation}\label{valuegradient}
    0= -\Rii u^{i}_{t} + \rii + B^{i\top}_{t}Z^{i}_{t+1}( \dynA x_{t} + \sum_{j=1}^{N}\dynB u^{j}_{t}) + B^{i\top}_{t}\zeta^{i}_{t+1}.
\end{equation}
\vspace{-0.4cm}\\
It is clear from \eqref{valuefunc} that the optimal $u^{i}_{t}$ is a function of state $x_{t}$, and \eqref{valuegradient} suggests an affine form, \( u^{i*}_{t} = -P^{i}_{t}x_{t} - \alpha^{i}_{t}\). Using this affine form, along with \eqref{lindyn} and \eqref{quadvalue}, yields two systems of equations \eqref{findpandalpha} which can be used to find the $P$'s and $\alpha$'s
\vspace{-0.3cm}
\begin{equation}\label{findpandalpha}
    \begin{split}(\Rii+B^{i\top}_{t}Z^{i}_{t+1}B^{i}_{t})P^{i}_{t}+B^{i\top}_{t}Z^{i}_{t+1}\sum_{j\neq i}\dynB P^{j}_{t} 
        &= B^{i\top}_{t}Z^{i}_{t+1}\dynA, \\
        (\Rii+B^{i\top}_{t}Z^{i}_{t+1}B^{i}_{t})\alpha^{i}_{t}+B^{i\top}_{t}Z^{i}_{t+1}\sum_{j\neq i}\dynB \alpha^{j}_{t}
        &= B^{i\top}_{t}\zeta^{i}_{t+1}+\rii.
    \end{split}
\end{equation}
\vspace{-0.4cm}\\
Ultimately, we substitute back into \eqref{valuefunc} and get the set of recursions (solved backwards in time)
\vspace{-0.3cm}
\begin{equation}\label{feedbackrecursions}
    \begin{split}
        Z^{i}_{t} &= Q^{i}_{t} + \sum_{j=1}^{N}P^{j\top}_{t}\controlR P^{j}_{t} + F^{\top}_{t}Z^{i}_{t+1}F_{t},\\
        \zeta^{i}_{t} &= q^{i}_{t} + \sum_{j=1}^{N}(P^{j\top}_{t}\controlR \alpha^{j}_{t} - P^{j\top}_{t}r^{ij}_{t})+F^{\top}_{t}(\zeta^{i}_{t+1} + Z^{i}_{t+1} \beta_{t}),\\
        n^{i}_{t} &= \frac{1}{2} \left[ \sum_{j=1}^{N}(\alpha^{j\top}_{t}\controlR - 2r^{ij\top}_{t})\alpha^{j}_{t}\;-\right.\left.(2\zeta^{i}_{t+1}-Z^{i}_{t+1} \sum_{j=1}^{N}B^{j}_{t}\alpha^{j}_{t})^{\top}\sum_{j=1}^{N}B^{j}_{t}\alpha^{j}_{t} \right] + n^{i}_{t+1},
    \end{split}
\end{equation}
\vspace{-0.3cm}\\
with \( F_{t} = \dynA - \sum_{j=1}^{N}\dynB P^{j}_{t} \), \( \beta_{t} = - \sum_{j=1}^{N}\dynB \alpha^{j}_{t} \) and the terminal conditions $Z^{i}_{T+1} = 0$, $\zeta^{i}_{T+1} = 0$ and $n^{i}_{T+1} = 0$.
\begin{rem}\label{remark: Feedback}
    Solving a feedback LQ game requires \(\{Z^i_{T+1},\zeta^i_{T+1}\}\) as inputs and finds $\{P^{i}_{t}, \alpha^{i}_{t}, Z^i_{t},\zeta^i_{t}\}_{t \in \mathbf{T}}~\forall~i\in \mathbf{N}$, as in Algorithm \ref{algo: solveLQFeedback}. The optimal control strategies for all players can then be found for any initial state \(x_1\) by using the affine-control form \( u^{i*}_{t} = -P^{i}_{t}x_{t} - \alpha^{i}_{t}\).
\end{rem}
\begin{algorithm}\label{algo: solveLQFeedback}
\SetKwComment{Comment}{/* }{ */}
\caption{Feedback LQ Game Solver (\texttt{solveLQFeedback})}
\KwIn{time horizon $T$, linear dynamics (\ref{lindyn}), costs (\ref{lqcost}), \(Z^i_{T+1}, \zeta^i_{T+1}~\forall~i\in\mathbf{N}\)}
\KwOut{$\{P^{i}_{t}, \alpha^{i}_{t}, Z^i_{t},\zeta^i_{t}\}_{t \in \mathbf{T}}~\forall~i\in \mathbf{N}$}
\For{time $t=T,\dots,1$}{
compute $P^i_t, \alpha^i_t~\forall~i\in\mathbf{N}$ through \eqref{findpandalpha} \hspace{1.6cm}\Comment{backward in time}
compute recursions given in \eqref{feedbackrecursions}
}
\KwRet{$\{P^{i}_{t}, \alpha^{i}_{t}, Z^i_{t},\zeta^i_{t}\}_{t \in \mathbf{T}}~\forall~i\in \mathbf{N}$}
\end{algorithm}
\begin{rem}
    For a feedback LQ game, \(Z^i_{t}\) and \(\zeta^i_{t}\) encode the optimal cost-to-go for each player \(i\) at time \(t\). In an open-loop LQ game, \(M^{i}_{t}\) and \(m^i_t\) play an analogous role.
\end{rem}
\noindent \textbf{Iterative LQ Games.}
The open-loop and feedback solutions presented above hold for dynamic games with time-varying linear dynamics and quadratic cost functions. To address games with nonlinear dynamics $x_{t} = f_{t}(x_{t}, u_{t}^{1:N})$, and nonquadratic costs for each player $i\in \mathbf{N}$ being sums of running costs $g^{i}_t(x_{t}, u^{1:N}_{t})$, we adopt the iterative linear-quadratic approximations framework for solving general-sum, $N$-person games \cite{fridovich2020efficient} which can be efficiently computed and is based on the iterative linear-quadratic regulator (ILQR) algorithm \cite{li2004iterative}. At each iteration, a full nonlinear system trajectory \(\{\hat{x}_{t}, \hat{u}^{1:N}_{t}\}_{t\in\mathbf{T}}\) is simulated. At each time $t$ and for each player $i$,  for arbitrary $x_{t}$ and $u^{i}_{t}$, consider the deviations from the trajectory iterate, \(\delta x_{t} = x_{t} - \hat{x}_{t}\) and \(\delta u^{i}_{t} = u^{i}_{t} - \hat{u}^{i}_{t}\). Then
\vspace{-0.3cm}
\begin{equation}\label{dynapprox}
    \delta x_{t+1} = A_{t}\delta x_t + \sum_{i=1}^{N}B^{i}_{t}\delta u^{i}_{t},
\end{equation}
\vspace{-0.5cm}\\
where $A_{t}$ and $B^{i}_{t}$ are the Jacobians $D_{\hat{x}}f_{t}(\hat{x}_{t}, \hat{u}^{1:N}_{t})$ and $D_{\hat{u}^{i}}f_{t}(\hat{x}_{t}, \hat{u}^{1:N}_{t})$ respectively. We assume no mixed partials, and take the quadratic approximations of running costs for each player $i$ as
\vspace{-0.415cm}
\begin{equation}\label{costquadapproxilq}
        g^{i}_t(x_{t}, u^{1:N}_{t}) \approx g^{i}_t(\hat{x}_{t}, \hat{u}^{1:N}_{t}) + \frac{1}{2}\delta x^{\top}_{t}(Q^{i}_{t}\delta x_{t}+2l^{i}_{t})+\frac{1}{2}\sum_{j=1}^{N} \delta u^{j\top}_{t}(R^{ij}_{t} \delta u^{j}_{t} + 2r^{ij}_{t}),
\end{equation}
\vspace{-0.5cm}\\
where $l^{i}_{t}$, $r^{ij}_{t}$ are the gradients $D_{\hat{x}}g^{i}_t$ and $D_{\hat{u}^{j}}g^{i}_t$ respectively; and $Q^{i}_{t}$, $R^{ij}_{t}$ are the Hessians $D^{2}_{\hat{x}\hat{x}}g^{i}_t$ and $D^{2}_{\hat{u}^{j}\hat{u}^{j}}g^{i}_t$ respectively. Using these approximations at each iteration, an LQ game of an appropriate information structure is solved to obtain strategies utilized to generate the next trajectory iterate. This is repeated until convergence, yielding a local approximation to a Nash equilibrium.




\vspace{-0.3cm}
\section{Planning with Occlusions: Notational Conventions}\label{Problem Statement}
\begin{figure}
    \centering
    \includegraphics[width=\columnwidth]{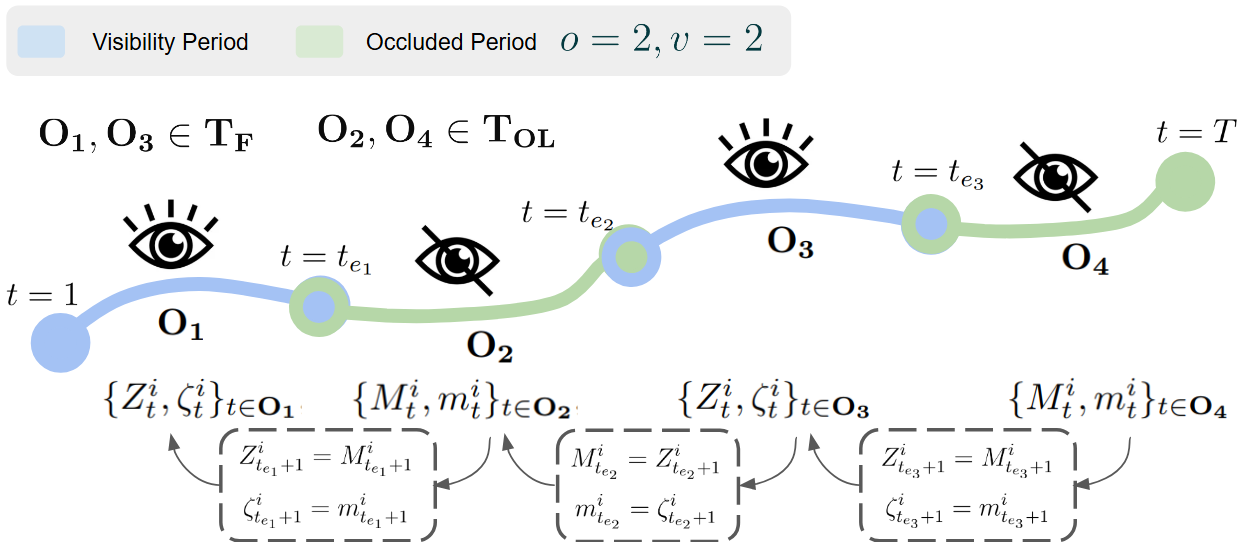}
    \caption{Illustration of the novel \emph{hybrid} information structure which we introduce in Algorithm \ref{algo: solveLQHybrid}. By taking terminal costs for each period from the cost-to-go of the next period, successive periods are linked together, which enables periods to reason about future occlusions. This figure represents a trajectory with two periods of occlusions and two periods of visibility \((o=2, v=2)\).}
    \label{algo_fig}
    \vspace*{-0.3cm} 
\end{figure}

We consider an $N$-player finite horizon general-sum dynamic game with nonlinear dynamics \(x_{t+1} = f_t(x_{t},u_{t}^{1:N})\) and nonquadratic costs \(J^i(u^1,\dots, u^N) = \sum_{t=1}^{T}g^{i}_t(x_{t},u^{1:N}_{t})\). We assume that at any time, players can determine whether they are occluded from other players or not. Critically, these occlusions are assumed to be such that at some time \(t\), either all players are mutually occluded or all players are mutually visible to each other. For $o$ periods of occlusions and $v$ periods of visibility throughout the game, consider the set of all time $\{1,\dots,T\}\equiv\mathbf{T}$ and partitions $\{\mathbf{O}_j | 1\leq j\leq o+v\}$, with each $\mathbf{O}_j$ being an interval of time indices within T denoting either a period of visibility or occlusion. Occlusions influence the information which each player $i$ has access to at any time step in the following way: at times $t\in \cup_{j=1}^{v}\mathbf{O}_{l_{j}}\equiv\mathbf{T_{F}}\subseteq\mathbf{T}$ when players are visible to each other, each player has access to the full game state, i.e., $u^{i}_t = \gamma_{t}^{i}(x_{t})$ for some functions $\gamma^{i}_t:\mathbb{R}^n\rightarrow\mathbb{R}^{m_{i}}, t \in \mathbf{T_{F}}$. For other times $t\in \cup_{j=1}^{o}\mathbf{O}_{n_{j}}\equiv\mathbf{T_{OL}} \equiv \mathbf{T}\setminus \mathbf{T_{F}} $ when one of $o$ occlusions restrict player visibility, players only know the state corresponding to the time $t_{o_{n_{j}}}$ when the $n_{j}^{th}$ occlusion \emph{started}, i.e., $u^{i}_t = \gamma^{i}_{t}(x_{t_{o_{n_{j}}}}), t \in \mathbf{O}_{n_j}, 1\leq j \leq o$. The time $t_{e_{n_{j}}}$ denotes the end of a period \(\mathbf{O}_{n_j}\). An example of this notation is in Fig. \ref{algo_fig}, denoting a hybrid information LQ game which will be solved in \Cref{HybridLQApproach}. As occlusion occurrences depend on the state trajectory and are unknown \emph{a priori}, $v$ and $o$ are unknown. Thus, while employing a hybrid information structure, players must determine whether an occlusion is present or not, and \emph{switch} between the two modes without being specifically instructed when to do so. 
\section{A Hybrid Information Structure Game Solver}\label{Our Approach}
We introduce our main contributions in this section. First, we develop a method that seeks Nash equilibrium solutions for a hybrid information linear quadratic (LQ) game, which involves switching between open-loop and feedback information structures during the game, assuming that occlusion patterns are \emph{known} beforehand to all players. Second, we employ this hybrid LQ game iteratively in our overall framework to solve a game with nonlinear dynamics, generalized costs, and a priori \emph{unknown} occlusion patterns.

\vspace{-0.3cm}
\subsection{Solving a Hybrid Information LQ Nash Game}\label{HybridLQApproach}
We first solve a general-sum, $N$-player dynamic game with linear, time-varying dynamics, quadratic costs, a finite horizon of $T$ time steps, and all $\mathbf{O_{j}}$ \emph{specified}.\\
\textbf{Hybrid Approach.} We use the feedback and open-loop solutions of LQ games described in Section \ref{Preliminaries} at different time steps in the game. This game is solved backwards in time, as in Algorithm \ref{algo: solveLQHybrid}. For each period $\mathbf{O_{j}}$, we employ a feedback solution if $\mathbf{O_{j}\in T_{F}}$ or an open-loop solution if $\mathbf{O_{j}\in T_{OL}}$. From \eqref{findpandalpha}, the feedback portion of the game yields $\{P^{i}_{t}, \alpha^{i}_{t}\}~\forall~i\in \mathbf{N},~\forall~t \in \mathbf{T_{F}}$, which we use to construct each players' affine control strategies, i.e., \(u^{i*}_{t} = -P^{i}_{t}x_{t} - \alpha^{i}_{t} \; \forall \; i \in \mathbf{N}\). For other portions, the open-loop solutions are found by \eqref{reclambda}, which yields $\{M^{i}_{t}, m^i_{t}\}, \forall \; i \in \mathbf{N}, \forall\;t\in\mathbf{T_{OL}}$, which we use to find open-loop controls $\{u^{i*}_{t}\}_{t\in\mathbf{T_{OL}}} \; \forall \, i \in \mathbf{N}$ as in \eqref{openloopcontrols}.\\
\textbf{Challenge for Hybrid Approach.} Consider the times $t\in\{t_{e_{j}}\}_{1\leq j \leq o+v}$ when a period $\mathbf{O_{j}}$ ends. From \eqref{remark: OL} and \eqref{remark: Feedback}, solving successive open-loop and feedback LQ games in a hybrid manner as above requires \(\{M^{i}_{t_{e_{j}}}, m^{i}_{t_{e_{j}}}\}\) and \(\{ Z^{i}_{t_{e_{j}}+1}, \zeta^{i}_{t_{e_{j}}+1} \}\) respectively. However, because \(t_{e_j}\) will not be the terminal time \(T\) when \(j\neq o+v\), \(M^{i}_{t_{e_{j}}}, m^{i}_{t_{e_{j}}}, Z^{i}_{t_{e_{j}}+1},\) and \(\zeta^{i}_{t_{e_{j}}+1}\) cannot be chosen according to \eqref{reclambda} or \eqref{feedbackrecursions}. This challenge prohibits the \emph{as is} use of "pure" open-loop or feedback games for their respective periods.\\
\textbf{Linking Occluded and Visibility Periods.} To overcome the above challenge, Algorithm \ref{algo: solveLQHybrid} employs a novel \emph{switching} at times $t\in\{t_{e_{j}}\}_{1\leq j \leq o+v}$. The terminal cost-to-go required for solving the game for $\mathbf{O_{j}}$ are provided by information from $\mathbf{O_{j+1}}$. This links immediate periods of occlusion and visibility (or vice-versa) together (see Fig. \ref{algo_fig}). Thus, if $\mathbf{O_{j}\in T_{OL}}$, the terminal $\{M^{i}_{t_{e_{j}}}, m^{i}_{t_{e_{j}}}\}$ values of the open-loop game for $\mathbf{O_j}$ are taken as the optimal costs to go $\{ Z^{i}_{t_{e_{j}}+1}, \zeta^{i}_{t_{e_{j}}+1} \}$ from $\mathbf{O_{j+1}}$, already calculated in \eqref{feedbackrecursions}. Similarly, if $\mathbf{O_{j}\in T_{F}}$, the terminal $\{ Z^{i}_{t_{e_{j}}}, \zeta^{i}_{t_{e_{j}}} \}$ values of the feedback game for $\mathbf{O_j}$ are taken as $\{M^{i}_{t_{e_{j}}+1}, m^{i}_{t_{e_{j}}+1}\}$ from $\mathbf{O_{j+1}}$, available through \eqref{reclambda}.\par 
This switching not only allows successive open-loop and feedback portions to be solved but also provides meaningful linking between occluded and visible periods. For example, consider the overtaking example in Fig. \ref{intro_fig}. Here, when transitioning from occlusion to visibility, providing \(\{Z^{i}_{t_{e_{j}}+1}, \zeta^{i}_{t_{e_{j}}+1}\}\) denotes the optimal cost-to-go representing the period when the cars finally start seeing each other. Finally, the full trajectory is unrolled using the dynamics, and the algorithm returns the strategies and corresponding unrolled trajectory.
\begin{rem}
    \textbf{Computational Complexity. }As mentioned in \Cref{Related Work}, previous work with hybrid information structures scale exponentially. In comparison, Algorithm \ref{algo: solveLQHybrid} only scales polynomially (cubic) with respect to the state dimension. This can be verified through \cite[Corollary~6.1]{bacsar1998dynamic}.
\end{rem}

\RestyleAlgo{ruled}
\begin{algorithm}\label{algo: solveLQHybrid}
\SetKwFunction{Union}{Union}\SetKwFunction{solveLQFeedback}{solveLQFeedback}
\SetKwFunction{Union}{Union}\SetKwFunction{solveLQOpenloop}{solveLQOpenloop}
\SetKwFunction{Union}{Union}\SetKwFunction{UnrollTrajectory}{UnrollTrajectory}
\SetKwFunction{Union}{Union}\SetKwFunction{getTrajectory}{getTrajectory}
\SetKwComment{Comment}{/* }{ */}
\caption{Hybrid LQ Solver (\texttt{HybridLQGame})}

\KwIn{initial state $x_{1}$, linear dynamics (\ref{lindyn}), costs (\ref{lqcost}), time horizon $T$, $\mathbf{T_{OL}}$, $\mathbf{T_{F}}$, $\{\mathbf{O_{j}}\}_{j=1}^{o+v}$}
\KwOut{Strategies $\{\gamma^{1:N}_{t}\}_{t\in\mathbf{T}}$, Trajectory $\{x_{t}, u^{1:N}_{t}\}_{t\in\mathbf{T}}$}
\For{iteration $j=o+v,\dots,1$}{
\If{$\mathbf{O_{j}\in T_{F}}$}{
$\{P^{i}_{t}, \alpha^{i}_{t},Z^{i}_{t}, \zeta^{i}_{t}\}_{t\in\mathbf{S_j}}\gets$ \solveLQFeedback{$\mathbf{O_j}$, $\{A_{t}, B^{i}_{t}, l^{i}_{t}, Q^{i}_{t}, r^{ij}_{t}, R^{ij}_{t}\}_{t\in\mathbf{O_j}}, M^{i}_{t_{e_j}+1}, m^{i}_{t_{e_j}+1}$} \hspace{0.6cm} \Comment{from \eqref{feedbackrecursions}}} 
\If{$\mathbf{O_{j}\in T_{OL}}$}{
$\{M^{i}_{t}, m^{i}_{t}\}_{t\in\mathbf{U_j}}\gets$ \solveLQOpenloop{$\mathbf{O_j}$, $\{A_{t}, B^{i}_{t}, l^{i}_{t}, Q^{i}_{t}, r^{ij}_{t}, R^{ij}_{t}\}_{t\in\mathbf{O_j}}, Z^{i}_{t_{e_j}+1}, \zeta^{i}_{t_{e_j}+1}$} \hspace{0.8cm} \Comment{from \eqref{reclambda}}
}
}
\KwRet{$\{M^i_t, m^i_t\}_{t\in\mathbf{T_{OL}}},\{P^i_t,\alpha^i_t\}_{t\in\mathbf{T_{F}}}$}
\end{algorithm}
\subsection{Iterative Hybrid Game Algorithm for non-LQ Games}
We employ the solution to the game described in Section \ref{HybridLQApproach} in our overall \textbf{O}cclusion-aware \textbf{G}ame \textbf{Solve}r (\texttt{OGSolve}) in Algorithm \ref{algo: Iterative Hybrid Solve} in order to solve a game game in which agents’ ability to observe one another is state-dependent, as in the occluded planning example of Fig. \ref{intro_fig}. We begin with unrolling a trajectory iterate \(\xi^{k} \equiv 
\{\hat{x}_{t},\hat{u}_{t}^{1:N}\}_{t\in\mathbf{T}}\) using an initial state $x_{1}$ and initial control strategies $\{u^{i^1}_{t}\}_{i\in\mathbf{N}}$ according to the dynamics \(f_t\). Similar to the iterative LQ method described in Section \ref{Preliminaries}, we define deviations from the trajectory iterate, \(\delta x_{t}\) and \(\delta u^{i}_{t}\). These are used to compute discrete time, linear approximations to the dynamics about $\xi^{k}$, as well as quadratic approximations to the running costs about $\xi^{k}$ as in \eqref{dynapprox} and \eqref{costquadapproxilq} respectively. Further, the players use the current trajectory iterate $\xi^{k}$ to determine occlusions in this trajectory, finding $o$, $v$ and the sets $\mathbf{O_{j}}, j \in \{1,2,\dots, o+v\}$ specific to $\xi^{k}$. This leads to a discrete-time LQ game, which is solved using Algorithm \ref{algo: solveLQHybrid}, and results are used to create candidate strategies $\delta \tilde{u}^{k}_{i}$. To avoid divergence caused by trajectories resulting from candidate strategies moving too far away from $\xi^k$ for dynamics linearizations and cost quadraticizations to hold with high accuracy, we only take a small step in the direction of $\delta\tilde{u}^{k}_{i}$ to obtain the next strategy set and generate the next trajectory iterate. This procedure is repeated until convergence, resulting in an approximation of the Nash equilibria that we seek.\\
Algorithm \ref{algo: Iterative Hybrid Solve} makes use of the following functions in its iterative framework:
\begin{itemize}
    \item \texttt{getTrajectory:} unrolls a trajectory for some given initial state and control strategies according to dynamics \(f_t\).
    \item \texttt{linearizeDynamics:} finds linear approximations of dynamics \(f_t\) along a given trajectory according to \eqref{dynapprox}.
    \item \texttt{quadraticCost:} finds quadratic approximations to cost \(g^i_t\) along a given trajectory according to \eqref{costquadapproxilq}.
    \item \texttt{findOcclusions:} determines the times \(\mathbf{T_{OL}}\) and \(\mathbf{T_{F}}\) when players were occluded and visible to each other, respectively, for a given trajectory. This is discussed in \Cref{experiments}.
    \item \texttt{getStrategy:} 
    finds controls \(u^{i*}_t~\forall~i \in \mathbf{N}, t \in \mathbf{T}\) for given \(\{M^i_t, m^i_t\}_{t\in\mathbf{T_{OL}}}^{i\in\mathbf{N}},\) \( \{P^i_t, \alpha^i_t\}_{t\in\mathbf{T_{F}}}^{i\in\mathbf{N}} \) according to \eqref{remark: OL} and \eqref{remark: Feedback}.
    \item \texttt{stepToward:} Computes \(u_t^{i^{k+1}} = u_t^{i^k} + \eta_t \delta \tilde{u}^{i^{k}}_{t}~\forall~i \in \mathbf{N}, t \in \mathbf{T}\) for an appropriate step size \(\eta_t\in(0,1]\). 
    
\end{itemize}

\begin{algorithm}\label{algo: Iterative Hybrid Solve}
\SetKwFunction{Union}{Union}\SetKwFunction{quadraticizeCosts}{quadraticCost}
\SetKwFunction{Union}{Union}\SetKwFunction{linearizeDynamics}{linearizeDynamics}
\SetKwFunction{Union}{Union}\SetKwFunction{getTrajectory}{getTrajectory}
\SetKwFunction{Union}{Union}\SetKwFunction{getStrategy}{getStrategy}
\SetKwFunction{Union}{Union}\SetKwFunction{solveLQHybrid}{solveLQHybrid}
\SetKwFunction{Union}{Union}\SetKwFunction{findOcclusions}{findOcclusions}
\SetKwFunction{Union}{Union}\SetKwFunction{stepToward}{stepToward}
\SetKwComment{Comment}{/* }{ */}
\caption{Occlusion Aware Game Solver (\texttt{OGSolve})}

\KwIn{initial state $x_1$ and control strategies $\{u^{i^1}_{t}\}_{t\in\mathbf{T}}^{i\in\mathbf{N}}$, costs $\{g^{i}_t\}_{t\in\mathbf{T}}^{i\in\mathbf{N}}$, time horizon $T$, dynamics $f$, constants $0<\eta_t\leq 1, t\in\mathbf{T}$}
\KwOut{Converged Trajectory $\{x_{t}, u^{1:N}_{t}\}_{t\in\mathbf{T}}$}
\For{iteration $k=1,2,\dots$}{
$\xi^k \leftarrow$ \getTrajectory{$x_{1}, f, \{u^{i^{k-1}}_{t}\}_{t\in\mathbf{T}}^{i\in\mathbf{N}}$} \hspace{2.2cm}  
$\texttt{Dyn}_k\equiv\{A_{t}, B^{i}_{t}\}_{t\in\mathbf{T}}^{i\in\mathbf{N}}\leftarrow$ \linearizeDynamics{$\xi^{k}, f$} 
$\texttt{Costs}_k\equiv\{l^{i}_{t}, Q^{i}_{t}, r^{ij}_{t}, R^{ij}_{t}\}_{t\in\mathbf{T}}^{i,j\in\mathbf{N}}\leftarrow$ \quadraticizeCosts{$\xi^k, g^i_t$} 
$\mathbf{T_{OL}^\mathnormal{k}}, \mathbf{T_{F}^\mathnormal{k}}, \{\mathbf{O_{j}^{\mathnormal{k}}}\}_{j=1}^{o_k+v_k}\leftarrow$ \findOcclusions{$\xi^k$} \\
$\{M^i_t, m^i_t\}^{i\in\mathbf{N}}_{t\in\mathbf{T_{OL}^\mathnormal{k}}},\{P^i_t,\alpha^i_t\}^{i\in\mathbf{N}}_{t\in\mathbf{T_{F}^\mathnormal{k}}}$ $\leftarrow$\texttt{HybridLQGame(}$x_1, \texttt{Dyn}_k, \texttt{Costs}_k, T, \mathbf{T_{OL}^\mathnormal{k}}, \mathbf{T_{F}^\mathnormal{k}}, \{\mathbf{O_{j}^{\mathnormal{k}}}\}_{j=1}^{o_k+v_k}$\texttt{)}\\
$\{\delta \tilde{u}^{i^k}_{t}\}_{t\in\mathbf{T}}^{i\in\mathbf{N}}\leftarrow$ \getStrategy{ 
 $\{M^i_t, m^i_t\}_{t\in\mathbf{T_{OL}^\mathnormal{k}}}^{i\in\mathbf{N}},\{P^i_t,\alpha^i_t\}_{t\in\mathbf{T_{F}^\mathnormal{k}}}^{i\in\mathbf{N}}$}\\
$\{u^{i^{k+1}}_{t}\}_{t\in\mathbf{T}}^{i\in\mathbf{N}}\leftarrow$\stepToward{$\{\delta \tilde{u}^{i^{k}}_{t}, \alpha_t\}_{t\in\mathbf{T}}^{i\in\mathbf{N}}, \xi^{k}$}\\
\If{converged}{\KwRet{\getTrajectory{$x_{1}, f, \{u^{i^{k}}_{t}\}_{t\in\mathbf{T}}^{i\in\mathbf{N}}$}}}
}
\end{algorithm}

\section{Experiments}\label{experiments}
We implement Algorithm \ref{algo: Iterative Hybrid Solve} in Julia\footnote{We will make our code open source upon publication.} and consider two driving simulation scenarios -- (i) three-player overtaking and (ii) two-player occluded intersection. We conduct experiments to validate the following \textbf{hypotheses} related to the hybrid information structure we propose in this paper and use in \texttt{OGSolve}:
\begin{enumerate}
    \item[\textbf{H1.}] The hybrid information structure is advantageous compared to an open-loop structure in occluded settings for motion planning.
    \item[\textbf{H2.}] Players employing a hybrid information structure can reason about safe motion planning even when beginning in occlusion.
\end{enumerate}
\subsection{Experiment Setup Details}
\begin{figure}
    \centering
    \includegraphics[width=0.4\columnwidth]{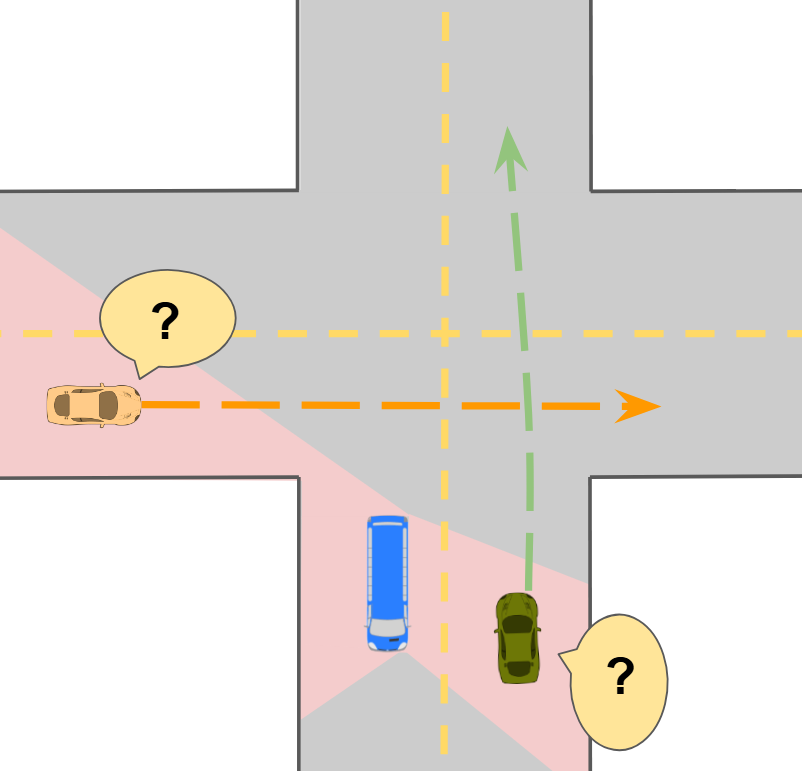}
    \caption{Traffic Intersection: Player 1 (green car) and Player 2 (orange car) are occluded due to a stationary blue bus at an intersection. Both want to keep going forward on their respective paths.}
    \label{exp_2_prob}
    \vspace*{-0.3cm} 
\end{figure}

For both overtaking and intersection scenarios, we play a game for $T=100$ time steps and a discretization of 0.1 s. All players are vehicles on the road, modelled as two-dimensional rectangles. We model each Player $i$'s motion using unicycle dynamics consisting of player rectangle centre positions, speed and heading. We consider a 4N-dimensional global state vector consisting of the concatenation of all player states, $x = (p_{i_{x}}, p_{i_{y}}, v, \theta_i)_{i=1}^{N}$ and 2-dimensional control input, $u_{i}:= (\dot{\theta}_i, \dot{v}_i),\; i = 1,\dots, N$. The running costs $g^i_t$ for each player are taken as weighted sums of the following: 
\begin{equation}\label{g}
    \begin{aligned}
    & \mathrm{Goal:} \; ||p_i-p_{goal, i}||^2, \\
    & \mathrm{Nominal \; Speed:}\; (v_i - v_{nom,i})^2, \\
    & \mathrm{Control \; input:} \; u^T_iR_{ii}u_i, \\
    & \mathrm{Distance \;from \;Lane \;Center:} \; d_{lane_i}(p_i)^2,\\
    & \mathrm{Lane\;Crossing\;Penalty:}\;\mathbf{1}\{d_{lane_i}(p_i)>d_{lane}\}(d_{lane}-d_{lane_i}(p_i))^2,\\
&\mathrm{Mutual\;Proximity:}\;\mathbf{1}\{||p_i-p_j||<d_{prox}\}(d_{prox} - ||p_i-p_j||)^2,\\
    &\mathrm{Speed\;Bounds:}\;\mathbf{1}\{v_i>\overline{v}_i\}(v_i-\overline{v}_i)^2 + \mathbf{1}\{v_i<\underline{v}_i\}(\underline{v}_i-v_i)^2,
    \end{aligned}
\end{equation}
\noindent where $p_i:=(p_{x_{i}}, p_{y_{i}})$ denotes player $i$'s position, $\mathbf{1}\{\cdot\}$ is the indicator function and $d_{lane_i}(p_i)$ denotes the distance of player $i$ from their lane center-line. $d_{lane}$ and $d_{prox}$ denote the lane half-width and threshold distance between players, and we set these values to 3.75m and 3m, respectively. The roads are assumed to have two lanes, one for each direction. The goal for each player is to reach the end of their lanes opposite to where they started, while the other costs encourage good driving behaviour.\\
\textbf{Choice of \texttt{findOcclusions}. }Let $\mathrm{Rect(i)}$ denote the two-dimensional rectangle representing Player \(i\). For some given state, we considered Players \(i\) and \(j\) to be visible to each other if it was possible to draw a line segment between some point in $\mathrm{Rect(i)}$ and some point in $\mathrm{Rect(j)}$ such that no occluding obstacle intersected this line. In pratice, one can use any off-the-shelf occlusion-checking algorithm incorporating field of view and point of view effects, robot sensing capabilities, etc. \texttt{OGSolve} is agnostic to how occlusions are found.\\ 
\textbf{Three-Player Overtaking. } This scenario consists of two cars and a truck, as in Fig. \ref{intro_fig}. Player 1 (green car) seeks to overtake Player 2 (blue truck) while maintaining sufficient distance from Player 3 (brown car), who comes in the opposite direction. Player 1 and Player 3 were modelled as having the same dimensions of \(4.48\)m\(\times1.76\)m, while Player 2 had the dimensions \(13.6\)m\(\times2.25\)m. Note that while we play a three-player game, in actuality, only Player 1 and Player 3 face occlusions. We choose \(g_t^2\) such that Player 2 is heavily penalised for changing its initial velocity to model the high inertia of a large truck.\\
\textbf{Two-Player Occluded Intersection. }We model a 2-player traffic intersection, as in Fig. \ref{exp_2_prob}. In this experiment, we set the initial conditions such that Players 1 and 2 are occluded till the intersection and will collide if they continue towards their target positions without 
 additional control input. Thus, this example represents a situation where the bulk of reasoning must be done in the occluded game portion. The running cost parameters are chosen such that Player 1 and Player 2 starts at the same speed but are asymmetrically penalised upon changing their speed, with Player 1 being more heavily penalised. 

\subsection{Results}
\begin{figure}
    \centering
    \includegraphics[width=0.6\columnwidth]{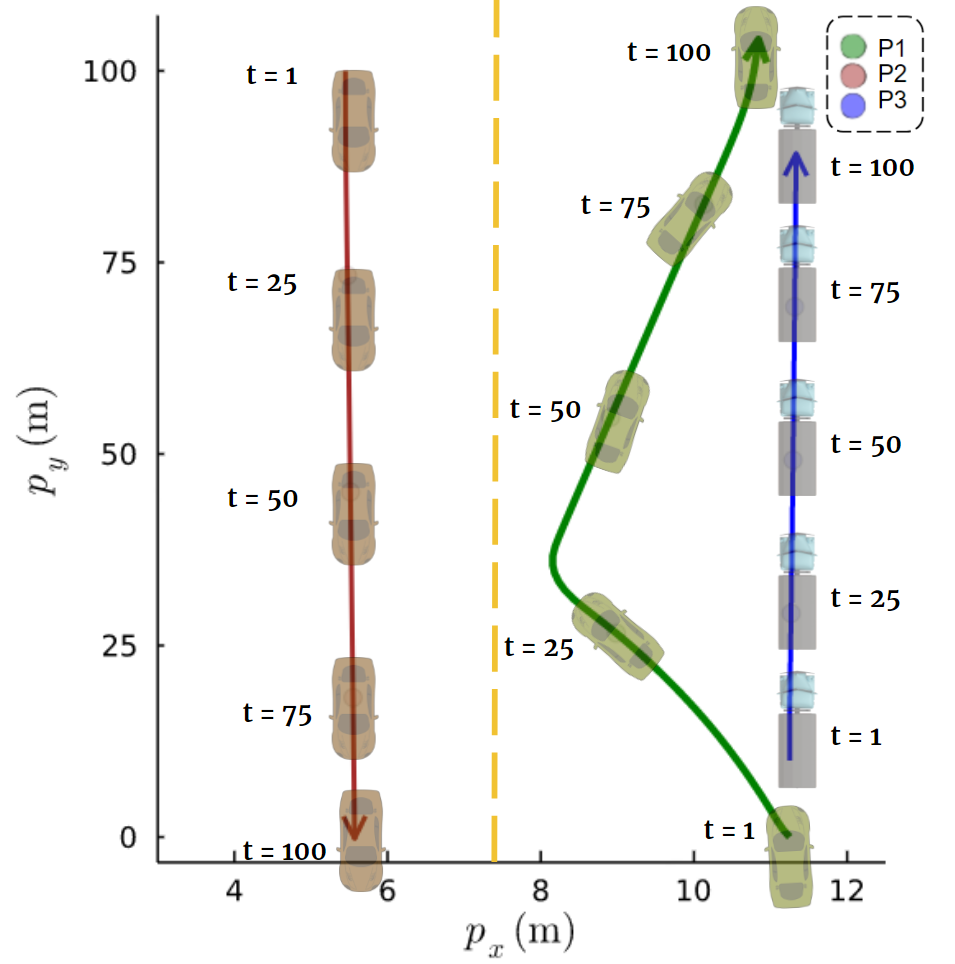}
    \caption{Three Player Overtaking Scenario: example of a converged trajectory obtained through \texttt{OGSolve}, with every $25^{th}$ time step marked. Players 1 \& 3 were found to be occluded for about 15\% of the time horizon.}
    \label{exp_1}
    \vspace*{-0.3cm} 
\end{figure}

\begin{figure}
    \centering
    \includegraphics[width=0.8\columnwidth]{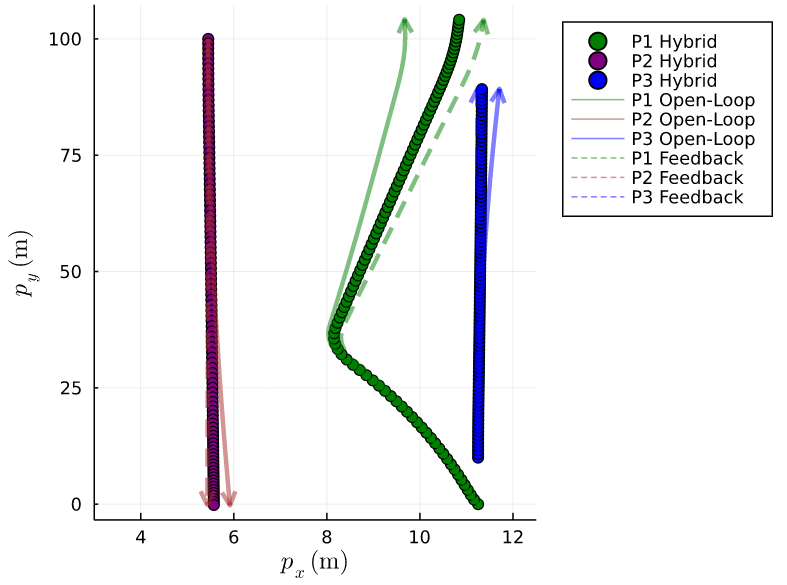}
    \caption{Hybrid information leads to better decision-making compared to the open-loop information structure. Comparison of \texttt{OGSolve} with pure feedback and pure open-loop solutions in the three-player overtaking scenario.}
    \label{exp_1_compare}
    \vspace*{-0.3cm} 
\end{figure}

\begin{figure}
    \centering
    \includegraphics[width=0.8\columnwidth]{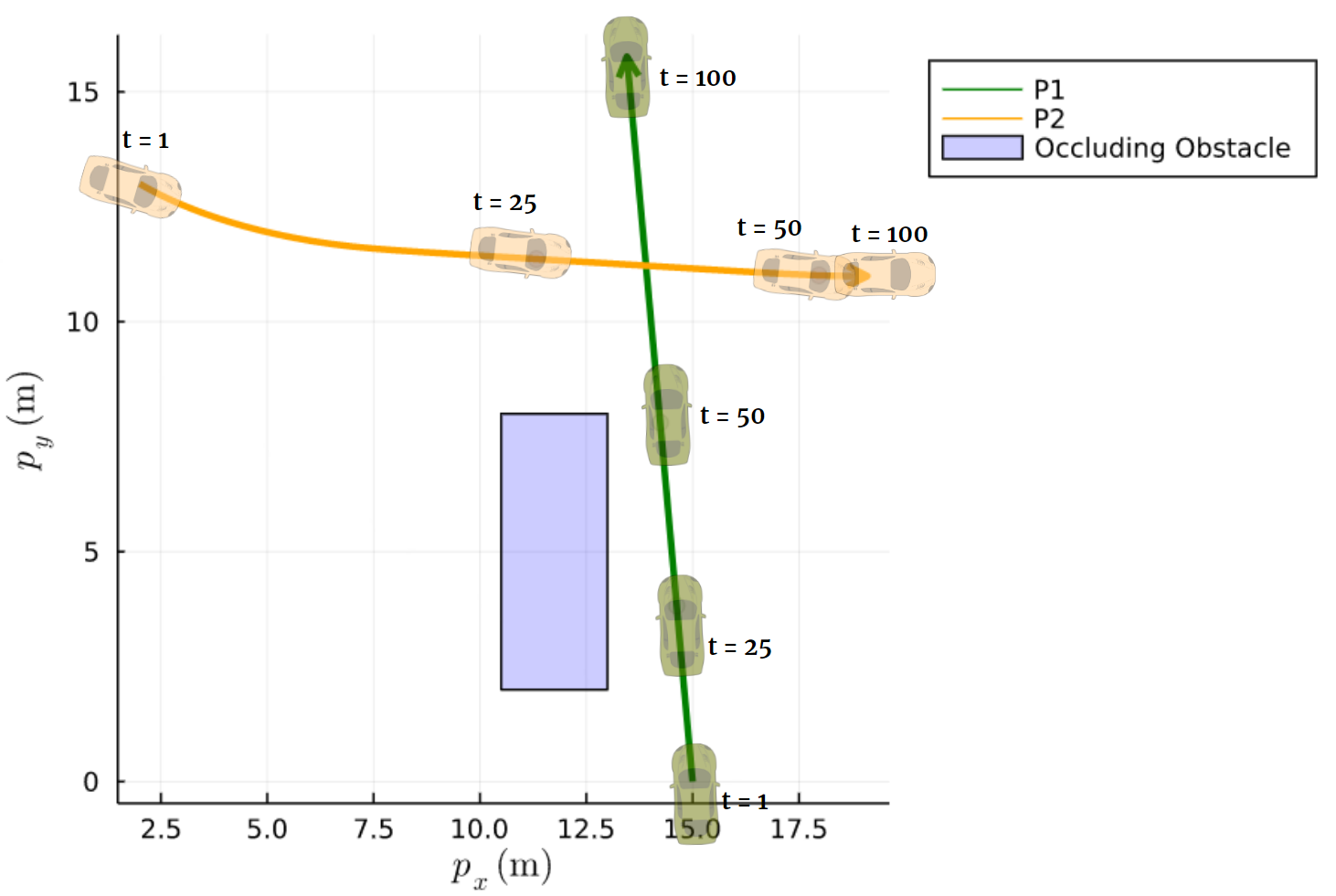}
    \caption{Results for the two-player intersection scenario.}
    \label{exp_2_res}
    \vspace*{-0.3cm} 
\end{figure}

\begin{figure}
    \centering
    \includegraphics[width=0.8\columnwidth]{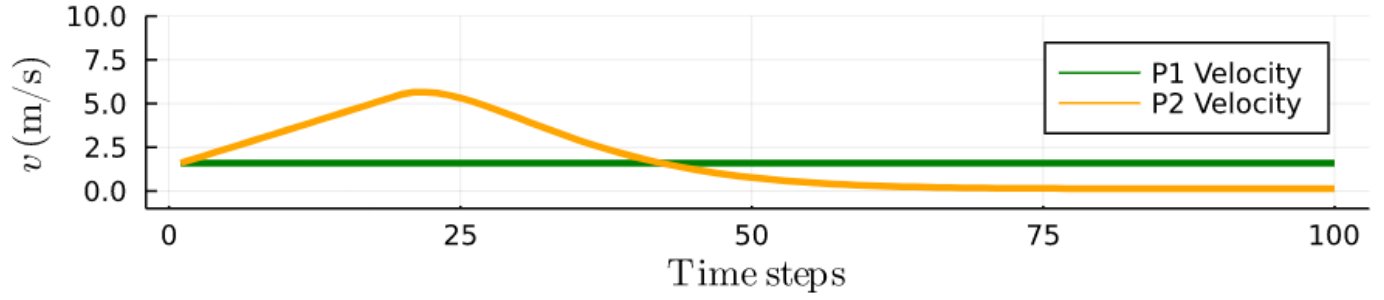}
    \caption{Player speed over time for the intersection scenario. Player 2 is able to zoom away from the intersection before slowing down. Player 1 is encouraged to keep its original speed, and only marginally decreases it's speed.}
    \label{exp_2_vel}
    \vspace*{-0.3cm} 
\end{figure}

Hypothesis \textbf{H1} is validated by analyzing the three-player overtaking scenario.\\ \\
\textbf{Hybrid information structure is advantageous compared to open-loop structure for motion planning.}
 The results for \texttt{OGSolve} for overtaking can be seen in Fig. \ref{exp_1} and Fig. \ref{exp_1_compare}. \texttt{OGSolve} was able to produce successful overtaking trajectories. Further, we compared \texttt{OGSolve} with pure feedback (assuming information ``leaked'' to all players through the occlusion) and pure open-loop strategies. The feedback trajectory was the most aggressive, while the open-loop trajectory was the most cautious. We interpret this as the feedback trajectory being more \emph{confident} than its open-loop counterpart due to the state being known to each player at all times in the feedback structure. \texttt{OGSolve} provided a convenient middle-ground and hybrid information, resulting in a more confident trajectory than the open-loop solution, better lane-keeping (see Fig. \ref{exp_1_compare}), and overall results in a safe trajectory. We benchmarked \texttt{OGSolve} by using the BenchmarkTools package~\cite{chen2016robust} in Julia on a laptop with a 2.3GHz Intel Core i7 CPU for 75 experiment runs\footnote{The number of experiment runs required for meaningful statistics is determined by BenchmarkTools. We did not choose the value of 75 ourselves.}. All runs had successful overtaking behaviour, taking a maximum of 170 iterations (\(k\) in Algorithm \ref{algo: Iterative Hybrid Solve}) to converge. A representative converged trajectory is shown in Fig. \ref{exp_1}. Player initializations were random for each run but chosen such that Players 1 and 3 always started out in an occluded state.
 \\
\noindent Hypothesis \textbf{H2} is validated by analyzing the two-player occluded intersection scenario. Note that the overtaking scenario also validates \textbf{H2}. \\ \\
\textbf{Players with hybrid information structure can reason about safe maneuvering even when starting in occlusions.}
The results for \texttt{OGSolve} for the intersection scenario can be seen in Fig. \ref{exp_2_res} and Fig. \ref{exp_2_vel}. The hybrid information algorithm was successful in making Player 2 reason how to avoid colliding with Player 1. Further, the asymmetric speed deviation penalisation caused Player 2 to vary their speed more prominently. Player 2 varied their speed to avoid getting too close to Player 1, speeding ahead to successfully cross the intersection before Player 1 reached it (see Fig. \ref{exp_2_vel}). For this example, BenchmarkTools took 94 experiment runs with all runs converging successfully. A representative converged trajectory is depicted in Fig. \ref{exp_2_res}. The maximum number of iterations a run took to converge was 25.

\section{Conclusion, Limitations, \& Future Work}
We present an algorithm (\texttt{OGSolve}) to locally approximate a Nash equilibrium in hybrid-information dynamic games, which model decision-making settings where agents do not receive the state information of other players for some portions of the game. We demonstrate the use of such hybrid information games for multi-agent motion planning in the presence of occlusions. Although agents cannot be modelled with feedback information structures in such settings, using the hybrid information structure that we introduce is advantageous compared to a pure open-loop structure. 
\\
\textbf{Limitations. } Our work makes two assumptions that warrant possible restrictions on its real-life applicability. First, the occlusion scenario we consider assumes that either \emph{all} players are mutually occluded or none are. In real life, occlusion patterns that violate this assumption can arise. Second, our work assumes that the number of agents is known to each agent beforehand. Further, we note that although our experiments demonstrate reliable convergence in specific scenarios, theoretical results in this direction remain of critical importance. \\
\textbf{Future Work. } Future work will consider extensions to more complex occlusion scenarios that necessitate interleaving open-loop and feedback information structures on a per-player basis rather than only over time, as in the present paper. Another direction would be to make the number of players unknown via a contingency game formation along the lines of \cite{peters2024contingency}.

\begin{credits}
\subsubsection{\ackname} D. Fridovich-Keil gratefully acknowledges guidance from Andrew Packard in 2018, regarding a closely related problem. This work was supported by the National Science Foundation under Grants 2211548 and 2336840.

\subsubsection{\discintname}
The authors have no competing interests.
\end{credits}
%
%
%
%
\renewcommand{\bibfont}{\normalfont\footnotesize}
{\renewcommand{\markboth}[2]{}\printbibliography}




\end{document}